
\input harvmac.tex
\Title{CTP/TAMU-33/92}{{A Multimonopole Solution in String Theory}
\footnote{$^\dagger$}{Work supported in part by NSF grant PHY-9106593.}}

\centerline{
Ramzi~R.~ Khuri\footnote{$^*$}{Supported by a World Laboratory Fellowship.}}
\bigskip\centerline{Center for Theoretical Physics}
\centerline{Texas A\&M University}\centerline{College Station, TX 77843}

\vskip .3in
A multimonopole solution in Yang-Mills field theory is obtained by
a modification of the 't Hooft ansatz for a four-dimensional instanton.
Although this solution has divergent action near each source, it
can be used to construct an exact finite action multimonopole solution of
heterotic string theory, in which the divergences from the
Yang-Mills sector are precisely cancelled by those from the gravity sector.

\Date{4/92}

\def\sqr#1#2{{\vbox{\hrule height.#2pt\hbox{\vrule width
.#2pt height#1pt \kern#1pt\vrule width.#2pt}\hrule height.#2pt}}}
\def\Box{\mathchoice\sqr64\sqr64\sqr{4.2}3\sqr33}

\def\met {g_{\mu\nu}}

\lref\reyone {S.--J.~Rey {\it Axionic String Instantons
and their Low-Energy Implications}, Proceedings, Tuscaloosa 1989,
Superstrings and particle theory, p.291.}

\lref\reytwo {S.--J.~Rey, Phys. Rev. {\bf D43} (1991) 526.}

\lref\abenone{I.~Antoniadis, C.~Bachas, J.~Ellis and D.~V.~Nanopoulos,
Phys. Lett. {\bf B211} (1988) 393.}

\lref\abentwo{I.~Antoniadis, C.~Bachas, J.~Ellis and D.~V.~Nanopoulos,
Nucl. Phys. {\bf B328} (1989) 117.}

\lref\mtone{R.~R.~Metsaev and A.~A.~Tseytlin, Phys. Lett.
{\bf B191} (1987) 354.}

\lref\mttwo{R.~R.~Metsaev and A.~A.~Tseytlin,
Nucl. Phys. {\bf B293} (1987) 385.}

\lref\cfmp{C.~G.~Callan, D.~Friedan, E.~J.~Martinec
and M.~J.~Perry, Nucl. Phys. {\bf B262} (1985) 593.}

\lref\ckp{C.~G.~Callan,
I.~R.~Klebanov and M.~J.~Perry, Nucl. Phys. {\bf B278} (1986) 78.}

\lref\love{C.~Lovelace, Phys. Lett. {\bf B135} (1984) 75.}

\lref\fridven{B.~E.~Fridling and A.~E.~M.~Van de Ven,
Nucl. Phys. {\bf B268} (1986) 719.}

\lref\gepwit{D.~Gepner and E.~Witten, Nucl. Phys. {\bf B278} (1986) 493.}

\lref\quartet{D.~J.~Gross,
J.~A.~Harvey, E.~J.~Martinec and R.~Rohm, Nucl. Phys. {\bf B267} (1986) 75.}

\lref\dine{M.~Dine, Lectures delivered at
TASI 1988, Brown University (1988) 653.}

\lref\brone{E.~A.~Bergshoeff and M.~de Roo, Nucl.
Phys. {\bf B328} (1989) 439.}

\lref\brtwo{E.~A.~Bergshoeff and M.~de Roo, Phys. Lett. {\bf B218} (1989) 210.}

\lref\chsone{C.~G.~Callan, J.~A.~Harvey and A.~Strominger, Nucl. Phys.
{\bf B359} (1991) 611.}

\lref\chstwo{C.~G.~Callan, J.~A.~Harvey and A.~Strominger, Nucl. Phys.
{\bf B367} (1991) 60.}

\lref\bpst{A.~A.~Belavin, A.~M.~Polyakov, A.~S.~Schwartz and Yu.~S.~Tyupkin,
Phys. Lett. {\bf B59} (1975) 85.}

\lref\thooft{G.~'t~Hooft, Nucl. Phys. {\bf B79} (1974) 276.}

\lref\hoofan{G.~'t~Hooft, Phys. Rev. Lett., {\bf 37} (1976) 8.}

\lref\wil{F.~Wilczek, in {\it Quark confinement and field theory},
Ed. D.~Stump and D.~Weingarten, John Wiley and Sons, New York (1977).}

\lref\cofa{E.~Corrigan and D.~B.~Fairlie, Phys. Lett. {\bf B67} (1977) 69.}

\lref\jackone{R.~Jackiw, C.~Nohl and C.~Rebbi, Phys. Rev. {\bf D15} (1977)
1642.}

\lref\jacktwo{R.~Jackiw, C.~Nohl and C.~Rebbi, in {\it Particles and
Fields}, Ed. David Boal and A.~N.~Kamal, Plenum Publishing Co., New York
(1978), p.199.}

\lref\rkinst{R.~R.~Khuri, Phys. Lett. {\bf B259} (1991) 261.}

\lref\rkscat{C.~G.~Callan and R.~R.~Khuri, Phys. Lett. {\bf B261} (1991) 363.}

\lref\rkmant{R.~R.~Khuri, {\it Manton Scattering of String Solitons}
PUPT-1270 (to appear in Nucl. Phys. {\bf B}).}

\lref\rkdg{R.~R.~Khuri, {\it Some Instanton Solutions in String
Theory} to appear in Proceedings of the XXth International Conference on
Differential Geometric Methods in Theoretical Physics, World Scientific,
October 1991.}

\lref\rkthes{R.~R.~Khuri, {\it Solitons and Instantons in String Theory},
 Princeton University Doctoral Thesis, August 1991.}

\lref\rksing{M.~J.~Duff, R.~R.~Khuri and J.~X.~Lu, {\it String and
Fivebrane Solitons: Singular or Non-singular?}, Texas A\&M preprint,
CTP/TAMU-89/91 (to appear in Nucl. Phys. {\bf B}).}

\lref\rkorb{R.~R.~Khuri and H.~S.~La, {\it Orbits of a String around a
Fivebrane}, Texas A\&M preprint, CTP/TAMU-95/91
(submitted to Phys. Rev. Lett.).}

\lref\rkmot{R.~R.~Khuri and H.~S.~La, {\it String Motion in Fivebrane
Geometry}, Texas A\&M preprint, CTP/TAMU-98/91 (submitted to Nucl. Phys. B).}

\lref\rkmonex{R.~R.~Khuri {\it A Heterotic Multimonopole Solution},
Texas A\&M preprint, CTP/TAMU-35/92.}

\lref\rkmono{R.~R.~Khuri {\it A Multimonopole Solution in
String Theory}, Texas A\&M preprint, CTP/TAMU-33/92.}

\lref\rkmscat{R.~R.~Khuri {\it Scattering of String Monopoles},
Texas A\&M preprint, CTP/TAMU-34/92.}

\lref\ginsp{P.~Ginsparg, Lectures delivered at
Les Houches summer session, June 28--August 5, 1988.}

\lref\swzw {W.~Boucher, D.~Friedan and A.~Kent, Phys. Lett.
{\bf B172} (1986) 316.}

\lref\dghrr{A.~Dabholkar, G.~Gibbons, J.~A.~Harvey and F.~Ruiz Ruiz,
Nucl. Phys. {\bf B340} (1990) 33.}

\lref\dabhar{A.~Dabholkar and J.~A.~Harvey,
Phys. Rev. Lett. {\bf 63} (1989) 478.}

\lref\prso{M.~K.~Prasad and C.~M.~Sommerfield, Phys. Rev. Lett. {\bf 35}
(1975) 760.}

\lref\jim{J.~A.~Harvey and J.~Liu, Phys. Lett. {\bf B268} (1991) 40.}

\lref\mantone{N.~S.~Manton, Nucl. Phys. {\bf B126} (1977) 525.}

\lref\manttwo{N.~S.~Manton, Phys. Lett. {\bf B110} (1982) 54.}

\lref\mantthree{N.~S.~Manton, Phys. Lett. {\bf B154} (1985) 397.}

\lref\atiyah{M.~F.~Atiyah and N.~J.~Hitchin, Phys. Lett. {\bf A107}
(1985) 21.}

\lref\atiyahbook{M.~F.~Atiyah and N.~J.~Hitchin, {\it The Geometry and
Dynamics of Magnetic Monopoles}, Princeton University Press, 1988.}

\lref\strom{A.~Strominger, Nucl. Phys. {\bf B343} (1990) 167.}

\lref\gsw{M.~B.~Green, J.~H.~Schwartz and E.~Witten,
{\it Superstring Theory} vol. 1, Cambridge University Press (1987).}

\lref\polch{J.~Polchinski, Phys. Lett. {\bf B209} (1988) 252.}

\lref\dfluone{M.~J.~Duff and J.~X.~Lu, Nucl. Phys. {\bf B354} (1991) 141.}

\lref\dflutwo{M.~J.~Duff and J.~X.~Lu, Nucl. Phys. {\bf B354} (1991) 129.}

\lref\dfluthree{M.~J.~Duff and J.~X.~Lu, Phys. Rev. Lett. {\bf 66}
(1991) 1402.}

\lref\dflufour{M.~J.~Duff and J.~X.~Lu, Nucl. Phys. {\bf B357} (1991)
534.}

\lref\dfstel{M.~J.~Duff and K.~S.~Stelle, Phys. Lett. {\bf B253} (1991)
113.}

\lref\ferone{R.~C.~Ferrell and D.~M.~Eardley, Phys. Rev. Lett. {\bf 59}
(1987) 1617.}

\lref\fertwo{R.~C.~Ferrell and D.~M.~Eardley, {\it Slowly Moving
Maximally Charged Black Holes} in Frontiers in Numerical Relativity,
Cambridge University Press, 1987.}

\lref\gh{G.~W.~Gibbons and S.~W.~Hawking, Phys. Rev. {\bf D15}
(1977) 2752.}

\lref\ghp{G.~W.~Gibbons, S.~W.~Hawking and M.~J.~Perry, Nucl. Phys. {\bf B318}
(1978) 141.}

\lref\briho{D.~Brill and G.~T.~Horowitz, Phys. Lett. {\bf B262} (1991)
437.}

\lref\gidone{S.~B.~Giddings and A.~Strominger, Nucl. Phys. {\bf B306}
(1988) 890.}

\lref\gidtwo{S.~B.~Giddings and A.~Strominger, Phys. Lett. {\bf B230}
(1989) 46.}

\lref\raj{R.~Rajaraman, {\it Solitons and Instantons}, North Holland,
1982.}

\lref\chsw{P.~Candelas, G.~T.~Horowitz, A.~Strominger and E.~Witten,
Nucl. Phys. {\bf B258} (1984) 46.}

\lref\bogo{E.~B.~Bogomolnyi, Sov. J. Nucl. Phys. {\bf 24} (1976) 449.}

\lref\cogo{E.~Corrigan and P.~Goddard, Comm. Math. Phys. {\bf 80} (1981)
575.}

\lref\wardone{R.~S.~Ward, Comm. Math. Phys. {\bf 79} (1981) 317.}

\lref\wardtwo{R.~S.~Ward, Comm. Math. Phys. {\bf 80} (1981) 563.}

\lref\wardthree{R.~S.~Ward, Phys. Lett. {\bf B158} (1985) 424.}

\lref\groper{D.~J.~Gross and M.~J.~Perry, Nucl. Phys. {\bf B226} (1983)
29.}

\lref\ash{{\it New Perspectives in Canonical Gravity}, ed. A.~Ashtekar,
Bibliopolis, 1988.}

\lref\lich{A.~Lichnerowicz, {\it Th\' eories Relativistes de la
Gravitation et de l'Electro-magnetisme}, (Masson, Paris 1955).}

\lref\goldstein{H.~Goldstein, {\it Classical Mechanics}, Addison-Wesley,
1981.}

\lref\dflufive{M.~J.~Duff and J.~X.~Lu, Class. Quant. Grav. {\bf 9}
(1992) 1.}

\lref\dflusix{M.~J.~Duff and J.~X.~Lu, Phys. Lett. {\bf B273} (1991)
409.}

\lref\hlp{J.~Hughes, J.~Liu and J.~Polchinski, Phys. Lett. {\bf B180}
(1986).}

\lref\town{P.~K.~Townsend, Phys. Lett. {\bf B202} (1988) 53.}

\lref\duff{M.~J.~Duff, Class. Quant. Grav. {\bf 5} (1988).}

\lref\rossi{P.~Rossi, Physics Reports, 86(6) 317-362.}

\lref\gksone{B.~Grossman, T.~W.~Kephart and J.~D.~Stasheff, Commun. Math.
Phys. {\bf 96} (1984) 431.}

\lref\gkstwo{B.~Grossman, T.~W.~Kephart and J.~D.~Stasheff, Commun. Math.
Phys. {\bf 100} (1985) 311.}

\lref\gksthree{B.~Grossman, T.~W.~Kephart and J.~D.~Stasheff, Phys. Lett.
{\bf B220} (1989) 431.}

\newsec{Introduction}

In recent work\refs{\rkinst\rkdg\dfluone\dflutwo\strom\chsone{--}\chstwo},
several classical solitonic solutions of string theory with instanton
structure have been presented. Exact solutions in bosonic\refs{\rkinst,\rkdg}
and heterotic\refs{\chsone,\chstwo}\ string theory have been obtained by
observing that a generalized curvature\refs{\mtone,\mttwo} combining the
metric and antisymmetric tensor possesses instanton structure\rkinst.

In this paper we present an exact multimonopole solution of heterotic
string theory. We first obtain an analogous mutlimonopole solution in
Yang-Mills field theory, whose action diverges near each source.
We show that the string theory solution, however, is finite as
a result of the cancellation of divergent terms from the gauge and
gravitational sectors.

In section 2 we derive a multimonopole solution of four-dimensional
YM + scalar field theory via a modification of the 't Hooft
ansatz\refs{\hoofan\wil\cofa\jackone{--}\jacktwo} for the Yang-Mills instanton.
This solution is not in the Prasad-Sommerfield\prso\ limit, but has the
topology of $Q=1$ monopole sources, saturates the Bogomoln'yi bound\bogo\ and
exhibits the far field behaviour of multimonopole sources. The action for this
solution, however, diverges near each source.

We use this solution in section 3 to construct an exact heterotic multimonopole
solution. This string solution has finite action, as the divergences coming
from the Yang-Mills sector are precisely cancelled by those from the
gravitational sector. The resultant action reduces to the tree-level form and
is easily calculated.

A more detailed version of this letter is presented in \rkmonex.

\newsec{Multimonopole Solution in Field Theory}

Consider the four-dimensional Euclidean action
\eqn\ymact{S=-{1\over 2g^2}\int d^4x {\rm Tr} G_{\mu\nu}G^{\mu\nu},
\qquad\qquad \mu,\nu =1,2,3,4.}
For gauge group $SU(2)$, the fields may be written as $A_\mu=(g/2i)
\sigma^a A_\mu^a$ and $G_{\mu\nu}=(g/2i)\sigma^a G_{\mu\nu}^a$\ \
(where $\sigma^a$, $a=1,2,3$ are the $2\times 2$ Pauli matrices).
The equation of motion derived from this action is solved by the
't Hooft ansatz\refs{\hoofan\wil\cofa\jackone{--}\jacktwo}
\eqn\hfanstz{A_\mu=i \overline{\Sigma}_{\mu\nu}\partial_\nu \ln f,}
where $\overline{\Sigma}_{\mu\nu}=\overline{\eta}^{i\mu\nu}(\sigma^i/2)$
for $i=1,2,3$, where
\eqn\hfeta{\eqalign{\overline{\eta}^{i\mu\nu}=-\overline{\eta}^{i\nu\mu}
&=\epsilon^{i\mu\nu},\qquad\qquad \mu,\nu=1,2,3,\cr
&=-\delta^{i\mu},\qquad\qquad \nu=4 \cr}}
and where $f^{-1}\Box\ f=0$. The ansatz for the anti-self-dual solution
is similar, with the $\delta$-term in \hfeta\ changing sign.

To obtain a multi-instanton solution, one solves for $f$ in the
four-dimensional space to obtain
\eqn\finst{f=1+\sum_{i=1}^N{\rho_i^2\over |\vec x - \vec a_i|^2},}
where $\rho_i^2$ is the instanton scale size and $\vec a_i$ the location in
four-space of the $i$th instanton.

We obtain a multimonopole solution by modifying this procedure as follows.
We single out a direction in the transverse four-space (say $x_4$) and assume
all fields are independent of this coordinate. Then the solution for $f$ can
be written as
\eqn\fmono{f=1+\sum_{i=1}^N{m_i\over |\vec x - \vec a_i|},}
where $m_i$ is the charge and $\vec a_i$ the location in
the three-space $(123)$ of the $i$th monopole. If we make the identification
$\Phi\equiv A_4$ (we loosely refer to this field as a Higgs field in
this paper, even though there is no apparent symmetry breaking mechanism),
then the Lagrangian density for the above ansatz can be rewritten as
\eqn\lgdn{\eqalign{G_{\mu\nu}^a G_{\mu\nu}^a =&G_{ij}^a G_{ij}^a +
2G_{k4}^a G_{k4}^a \cr =&G_{ij}^a G_{ij}^a + 2D_k \Phi^a D_k \Phi^a , \cr}}
which has the same form as the Lagrangian density for YM + massless scalar
field in three dimensions.

We now go to $3+1$ dimensions with the Lagrangian density (signature $(-+++)$)
\eqn\ymhlag{{\cal L}=-{1\over 4}G_{\mu\nu}^a G^{\mu\nu a} -{1\over 2}
D_\mu \Phi^a D^\mu \Phi^a,}
and show that the above multimonopole ansatz is a static solution with
$A_0^a=0$
and all time derivatives vanish. The equations of motion in this limit are
given by
\eqn\ymheqs{\eqalign{D_i G^{jia}&=g\epsilon^{abc}(D^j\Phi^b)\Phi^c,\cr
D_i D^i \Phi^a&=0.\cr}}
It is then straightforward to verify that the above equations are solved by
the modified 't Hooft ansatz
\eqn\monsol{\eqalign{\Phi^a&=\mp{1\over g}\delta^{ai}\partial_i \omega,\cr
A_k^a&=\epsilon^{akj}\partial_j \omega,\cr}}
where $\omega\equiv \ln f$. This solution represents a multimonopole
configuration with sources at $\vec a_i=1,2...N$. A simple observation of
far field and near field behaviour shows that this solution does not arise
in the Prasad-Sommerfield\prso\ limit. In particular, the fields are singular
near the sources and vanish as $r\to\infty$. This solution can be thought of
as a multi-line source instanton solution, each monopole being interpreted
as an ``instanton string''\rossi.

The topological charge of each source is easily computed
($\hat\Phi^a\equiv {\Phi^a/|\Phi|}$) to be
\eqn\topo{Q=\int d^3x k_0={1\over 8\pi}\int d^3x\epsilon_{ijk}\epsilon^{abc}
\partial_i\hat\Phi^a\partial_j\hat\Phi^b\partial_k\hat\Phi^c=1.}
The magnetic charge of each source is then given by $m_i=Q/g=1/g$.
It is also straightforward to show that the Bogomoln'yi\bogo\ bound
\eqn\bobo{G_{ij}^a=\epsilon_{ijk}D_k\Phi^a}
is saturated by this solution. Finally, it is easy to show that
the magnetic field $B_i={1\over 2}\epsilon_{ijk}F^{jk}$ (where
$F_{\mu\nu}\equiv \hat\Phi^a G_{\mu\nu}^a-(1/g)\epsilon^{abc}\hat\Phi^a
D_\mu \hat\Phi^b D_\nu \hat\Phi^c$ is the gauge-invariant electromagnetic
field tensor defined by 't Hooft\thooft) has the the far field limit behaviour
of a multimonopole configuration:
\eqn\bfield{B(\vec x)\to \sum_{i=1}^N {m_i(\vec x - \vec a_i)\over
|\vec x - \vec a_i|^3},\qquad {\rm as}\quad r\to \infty.}
As usual, the existence
of this static multimonopole solution owes to the cancellation of the
gauge and Higgs forces of exchange--the ``zero-force'' condition.

We have presented all the monopole properties of this solution.
Unfortunately, this solution as it stands has divergent
action near each source, and this singularity cannot be simply removed
by a unitary gauge transformation. This can be seen for a single source
by noting that as $r\to 0$, $A_k\to {1\over 2}\left(U^{-1}\partial_k U\right)$,
where $U$ is a unitary $2\times 2$ matrix. The expression in parentheses
represents a pure gauge, and there is no way to get around the $1/2$ factor in
attempting to ``gauge away'' the singularity\raj. The field theory
solution is therefore not very interesting physically. As we shall see in the
next section, however, we can obtain an analogous finite action solution in
string theory.

\newsec{Exact Heterotic Multimonopole Solution}

We use the above solution to construct an exact multimonopole solution of
heterotic string theory. The derivation of this solution closely parallels that
of the multi-instanton solution presented in \refs{\chsone,\chstwo}, but
in this case, the solution possesses three-dimensional (rather than
four-dimensional) spherical symmetry near each source. Again the reduction is
effected by singling out a direction in the transverse space.

The tree-level supersymmetric vacuum equations for the heterotic string are
given by
\eqn\suei{\delta\psi_M=\left(\nabla_M-{\textstyle {1\over 4}}H_{MAB}\Gamma^{AB}
\right)\epsilon=0,}
\eqn\sueii{\delta\lambda=\left(\Gamma^A\partial_A\phi-{\textstyle{1\over 6}}
H_{AMC}\Gamma^{ABC}\right)\epsilon=0,}
\eqn\sueiii{\delta\chi=F_{AB}\Gamma^{AB}\epsilon=0,}
where $\psi_M,\ \lambda$ and $\chi$ are the gravitino, dilatino and gaugino
fields. The Bianchi identity is given by
\eqn\bianchi{dH=\alpha' \left(\tr R\wedge R-{\textstyle{1\over 30}}\Tr
        F\wedge F\right).}

The $(9+1)$-dimensional Majorana-Weyl fermions decompose down to
chiral spinors according to $SO(9,1)\supset SO(5,1) \otimes SO(4)$ for
the $M^{9,1}\to M^{5,1}\times M^4$ decomposition.
Let $\mu,\nu,\lambda,\sigma=1,2,3,4$ and $a,b=0,5,6,7,8,9$. Then the ansatz
\eqn\anstz{\eqalign{\met&=e^{2\phi}\delta_{\mu\nu},\cr g_{ab}&=\eta_{ab},\cr
H_{\mu\nu\lambda}&=\pm\epsilon_{\mu\nu\lambda\sigma}\partial^\sigma\phi\cr}}
with constant chiral spinors $\epsilon_\pm$ solves the supersymmetry
equations with zero background fermi fields provided the YM gauge
field satisfies the instanton (anti)self-duality condition
\eqn\yminst{F_{\mu\nu}=\pm {1\over 2}\epsilon_{\mu\nu}{}^{\lambda\sigma}
F_{\lambda\sigma}.}
An exact solution is obtained as follows. Define a generalized connection by
\eqn\genc{\Omega^{AB}_{\pm M}=\omega^{AB}_M\pm H^{AB}_M}
embedded in an SU(2) subgroup of the gauge group, and equate it
to the gauge connection $A_\mu$\dine\ so that $dH=0$ and the corresponding
curvature $R(\Omega_{\pm})$ cancels against the Yang-Mills field strength $F$.
The crucial point is that for $e^{-2\phi}\Box\ e^{2\phi}=0$ with the
above ansatz, the curvature of the generalized connection can be written in the
covariant form\rkinst
\eqn\gencurv{\eqalign{R(\Omega_\pm)_{\mu\nu}^{mn}
=&\delta_{n\nu}\nabla_m\nabla_\mu\phi
- \delta_{n\mu}\nabla_m\nabla_\nu\phi + \delta_{m\mu}\nabla_n\nabla_\nu\phi
- \delta_{m\nu}\nabla_n\nabla_\mu\phi \cr
&\pm \epsilon_{\mu mn\alpha}\nabla_\alpha\nabla_\nu\phi
\mp \epsilon_{\nu mn\alpha}\nabla_\alpha\nabla_\mu\phi ,\cr}}
from which it easily follows that
\eqn\gcinst{R(\Omega_\pm)^{mn}_{\mu\nu}=
\mp\half\epsilon_{\mu\nu}^{\ \ \ \lambda\sigma}
R(\Omega_{\pm})_{\lambda\sigma}^{mn}.}
Thus we have a solution with the ansatz \anstz\ such that
\eqn\exsol{F_{\mu\nu}^{mn}=R(\Omega_{\pm})_{\mu\nu}^{mn},}
where both $F$ and $R$ are (anti)self-dual.
This solution becomes exact since $A_\mu=\Omega_{\pm\mu}$
implies that all the higher order corrections vanish\refs{\dine,\brone,\brtwo}.
The self-dual solution for the gauge connection is then given by the 't Hooft
ansatz
\eqn\hfanstz{A_\mu=i \overline{\Sigma}_{\mu\nu}\partial_\nu \ln f.}
A multimonopole is obtained as in the previous section by
\eqn\fdmono{f=e^{-2\phi_0}e^{2\phi}=1+\sum_{i=1}^N{m_i\over
|\vec x - \vec a_i|},}
where $m_i$ is the charge and $\vec a_i$ the location in
the three-space $(123)$ of the $i$th monopole. If we again identify the
Higgs field as $\Phi\equiv A_4$, then the gauge and Higgs fields may be simply
written in terms of the dilaton as
\eqn\stmono{\eqalign{\Phi^a&=-{2\over g}\delta^{ia}\partial_i\phi,\cr
A_k^a&=-{2\over g}\epsilon^{akj}\partial_j\phi\cr}}
for the self-dual solution. For the anti-self-dual solution, the Higgs
field simply changes sign. Here $g$ is the YM coupling constant. Note
that $\phi_0$ drops out in \stmono.

The above solution (with the gravitational fields obtained
directly from \anstz\ and \fdmono) represents an exact multimonopole
solution of heterotic string theory and has the same structure in
the four-dimensional transverse space as the multimonopole solution
of the YM + scalar field action of section 2. If we identify
the $(123)$ subspace of the transverse space as the space part of the
four-dimensional spacetime (with some toroidal compactification, similar
to that used in \jim) and take the timelike direction as the usual $X^0$,
then the monopole properties described in the previous section carry
over directly into the string solution.

The string action contains a term $-\alpha' F^2$ which also diverges
as in the field theory solution. However, this divergence
is precisely cancelled by the term $\alpha' R^2(\Omega_\pm)$ in the
$O(\alpha')$
action. This result follows from the exactness condition
$A_\mu=\Omega_{\pm\mu}$
which leads to $dH=0$ and the vanishing of all higher order corrections
in $\alpha'$. Another way of seeing this is to consider the higher order
corrections to the bosonic action shown in \refs{\brone,\brtwo}. All such
terms contain the tensor $T_{MNPQ}$, a generalized curvature incorporating
both $R(\Omega_\pm)$ and $F$. The ansatz is contructed precisely so that this
tensor vanishes identically\refs{\rkinst,\rkdg}. The action thus reduces to
its lowest order form and can be calculated directly for a multi-source
solution from the expressions for the massless fields in the gravity sector.

The divergences in the gravitational sector in heterotic string theory thus
serve to cancel the divergences stemming from the field theory solution. This
solution thus provides an interesting example of how this type of cancellation
can occur in string theory, and supports the promise of string theory as a
finite theory of quantum gravity. Another point of interest is that the string
solution represents a supersymmetric multimonopole solution coupled to gravity,
whose zero-force condition in the gravity sector (cancellation of
the attractive gravitational force and repulsive antisymmetric field force)
arises as a direct result of the zero-force condition in the gauge sector
(cancellation of gauge and Higgs forces of exchange) once the gauge connection
and generalized connection are identified.

We now calculate the mass of the heterotic multimonopole configuration.
Naively, the mass can be calculated from the tree-level action
\eqn\volaction{S=-{1\over 2\kappa^2}\int d^3x\sqrt{g} e^{-2\phi}\left( R
+ 4(\nabla\phi)^2 - {H^2\over 12}\right).}
There is one subtlety we must consider, however (see \rkmant).
{}From the term $\sqrt{g} e^{-2\phi} R$ in the integrand of the action,
the action density in \volaction\ contains double derivative terms of the
metric component fields. In general, one would like to work with an action
which depends only on the fields and their first derivatives. This
problem was solved in general relativity by Gibbons and Hawking\refs
{\gh,\ghp}, who added a surface term which precisely cancelled the
double derivative terms in the action in general relativity.
The addition of a surface term does not, of course, affect the equations
of motion.

It turns out that there is a relatively straightforward generalization
of the Gibbons-Hawking surface term (GHST) to string
theory\refs{\gidone,\briho}.
By antisymmetry, the axion field does not contribute to the GHST and
the surface term in this case can be written in the simple form
\eqn\stghst{S_{GHST}=-{1\over \kappa^2}\int_{\partial M}\left(e^{-2\phi}K-
K_0\right),}
where $\partial M$ is the surface boundary and $K$ and $K_0$ are the
traces of the fundamental form of the boundary surface embedded in the
metric $g$ and the Minkowskian metric $\eta$ respectively. The correct
effective action is thus obtained by adding the surface term of \stghst\
to the volume term of \volaction:
\eqn\trueaction{S=-{1\over 2\kappa^2}\left[\int d^3x
\sqrt{g} e^{-2\phi}
\left( R + 4(\nabla\phi)^2 - {H^2\over 12}\right)
+2\int_{\partial M}\left(e^{-2\phi}K-K_0\right)\right].}

By using the equations of motion, the volume term $S_V$ can
be written as a surface term (see \rkmant):
\eqn\svoltwo{S_V=-{1\over \kappa^2}\int_{\partial M} \hat
n\cdot\vec \nabla e^{-2\phi}.}
Note that $\sqrt{g}$ has been absorbed into the surface measure of
$\partial M$.
Since we have separability of sources in the limit of surfaces of
infinite radius, we may therefore compute $S_V$ for a single monopole
configuration in three-space
\eqn\simono{\eqalign{e^{2\phi}&=1+{m\over r},\cr
g_{ij}&=e^{2\phi}\delta_{ij},\cr}}
and simply add the contributions of an arbitrary number of sources.
The contribution of a single monopole to the static volume action is given by
\eqn\mvol{\eqalign{S_V&=-{1\over \kappa^2}
({\partial\over \partial r}e^{-2\phi})A(M)\cr
&=-{4\pi m\over \kappa^2}\cr}}
in the $r\to\infty$ limit,
where $A(M)=4\pi r^2(1+m/r)$ is the area of the boundary surface.

We now turn to the GHST.
A simple calculation of the extrinsic curvature $K$ for a single monopole
configuration \simono\ gives
\eqn\kcurved{K={2\over r^2}e^{-3\phi}(r+m/2).}
When the surface $\partial M$ is embedded in flat space, the radius of
curvature $R$ is given by $R=re^\phi$. The extrinsic curvature $K_0$ is
then given by
\eqn\kflat{K_0={2\over R}={2\over r}e^{-\phi}.}
The GHST is therefore  given by
\eqn\ghstone{S_{GHST}=-{2\over \kappa^2r}
\left(e^{-5\phi}( 1+ {m\over 2r}) - e^{-\phi}\right) A(M)
={12\pi m\over \kappa^2}}
in the $r\to\infty$ limit.

The total static action for a multi-soliton configuration, equal to the
total mass of the solitons, can then be obtained by adding the static
contributions to the action of the volume part and the GHST. The result is
\eqn\totmass{M_T={8\pi\over \kappa^2}\sum_{n=1}^N m_n.}
For our multimonopole configuration, however, it should be noted that
$m_n=1/g$ for $n=1,2...N$.

\newsec{Conclusion}

In this paper, we presented an exact multimonopole solution of
heterotic string theory. This solution represents a supersymmetric
extension of the bosonic string multimonopole solution outlined in
\rkthes, and is obtained by a modification of the 't Hooft
ansatz for a four-dimensional instanton. Exactness was shown by the
generalized curvature method used in \refs{\rkinst,\rkdg,\chsone,\chstwo}
to obtain exact instanton solutions in bosonic and heterotic string
theory. Unlike the instanton solutions, however, the monopole solutions
do not seem to be easily describable in terms of conformal field theories,
an unfortunate state of affairs from the point of view of string theory.

An analogous multimonopole solution of the four dimensional field theory
of YM + massless scalar field was first written down. This solution
possesses the properties of a multimonopole solution (topology,
far-field limit and Bogomoln'yi bound) but has divergent action near each
source. We used this solution to construct the string theory solution,
for which the divergences in the YM sector are cancelled by similar divergences
in the gravity sector, resulting in a finite action solution. This finding is
significant in that it represents an example of how string theory
incorporates gravity in such a way as to cancel infinities inherent in
gauge theories, thus supporting its promise as a finite theory of quantum
gravity.

\vfil\eject
\listrefs
\bye